\documentclass[aps,prl,reprint,a4paper,showpacs,notitlepage]{revtex4-1}%plr=PRL,notitlepage, showpacs
\usepackage{amsmath,amssymb}
\usepackage{graphicx}
\usepackage{color}	% required for `\textcolor' (yatex added)

\begin{document}

\title{Minimum vertex cover problems on random hypergraphs:\\
 replica symmetric solution and a leaf removal algorithm}
\date{\today}

\author{Satoshi Takabe}
\email{E-mail: s{\_}takabe@huku.c.u-tokyo.ac.jp}
\author{Koji Hukushima}
\affiliation{Graduate School of Arts and Sciences, The University of Tokyo, 3-8-1 Komaba, Meguro-ku, Tokyo 153-8902, Japan}

\begin{abstract}
We study minimum vertex cover problems on random $\alpha$-uniform hypergraphs
 using two different approaches, a replica method in statistical mechanics of random systems
 and a leaf removal algorithm.
 It is found that there exists a phase transition at the critical
 average degree $e/(\alpha-1)$. 
 Below the critical degree, a replica symmetric ansatz in the statistical-mechanical method holds
 and the algorithm estimates a solution of the problem which coincides with that by the replica method.
 In contrast, above the critical degree,
 the replica symmetric solution becomes unstable and these methods fail to estimate the exact solution.
 These results strongly suggest a close relation between the replica symmetry and
 the performance of approximation algorithm.
\end{abstract}


\maketitle

The more crucial part of everyday life computers bear, the more significance computer science
 and information theory seem to have.
 In particular, the computational complexity theory shows the difficulty,
 the limit of improving algorithms, to solve theoretical computational problems.
 It has revealed that the problems belong to several classes such as P and NP and there are many inclusion
 relations between these classes. For example, 2-satisfiability problems (2-SAT) belong to a class of P
 guaranteed to be solved in polynomial time.
 3-SAT and the vertex cover problems belong to a class of NP-complete~\cite{21}.
 %that is the other NP problems (we can confirm whether the instance
 %is a solution of a problem in polynomial time) can be reduced to them in polynomial time.
 These problems are deeply related to the well-known P versus NP problem plaguing
 the theoretical computer scientists, who have studied the worst-case performance to
 solve the computational problems.
 Among many types of combinatorial optimization problems, the minimum vertex cover problem  (min-VC) 
 belongs to a class of NP-hard. The approximation algorithm for the min-VC and its performance
 have been studied~\cite{27}. The application of the problem is to search a file on a file storage~\cite{pf}
 and to improve the group testing~\cite{gt}.

In addition to the worst-case analysis, an important alternative is the study of typical-case behavior
 on a class of random instances of the computational problems.
 Recently, statistical-mechanical methods of random spin systems have been applied to the problems
 such as $K$-SAT and constraint-satisfaction problems~\cite{26}.
 These methods, developed in the spin-glass theory~\cite{sg},
 enable us to study the typical properties of the randomized problems. For example, the statistical-mechanical
 approaches find a SAT/UNSAT transition of $K$-SAT~\cite{25},
 $p$-XOR-SAT~\cite{23}, $q$-coloring~\cite{zk} and min-VC~\cite{wh1,wh2,zh,3}.
% and a P/NP transition of $(2+p)$-SAT~\cite{22}
% with varying a parameter which characterizes the randomness.
% The min-VCs on random graphs are solved by a replica method~\cite{wh1,wh2}, a cavity method~\cite{zh},
% and a message passing~\cite{3}.
 These results clarify that there is a so-called replica symmetric (RS) phase
 where a replica symmetry ansatz provides correct estimates of the typical properties,
 and a replica symmetry breaking (RSB) phase where those estimates become unstable.
% These results clarify that there is a so-called replica symmetric (RS) phase
% where we can estimate the size of the min-VC, and a replica symmetry breaking (RSB) phase where the estimate
% becomes unstable.
 Together with these approaches, a typical-case performance of some
 approximation algorithms has been also studied~\cite{bg,24,dh}, suggesting
 that there is a non-trivial relation between the replica symmetry and the
 performance of approximation algorithms.

In this Letter, we study the minimum vertex cover problem on a random hypergraph.
 The random graph is defined by two distributions, the degree distribution and the edge size distribution.
 The degree means the number of edges connecting to a vertex and the edge size represents the number of vertices
 connected to an edge.
 As the former distribution, the Poisson distribution and the delta function are
 often used and they are called an Er\"odos-R\'enyi random graph and a regular random graph, respectively~\cite{26}.
 As the latter distribution, one uses the delta function with a mean $\alpha$, which yields a random graph with the
 same edge size as $\alpha$ called a random $\alpha$-uniform hypergraph.
 In general, a statistical-mechanical model defined on a hypergraph has multi-body interactions determined by its
 edge size.
% The issue of random graph theory is
% to clarify the property of random graphs and the point of difference between these types of graphs.
 In contrast to a conventional two-body interaction, the higher-order multi-body interactions often change a type of phase
 transition and a breaking pattern of the replica symmetry as shown in the $p$-body spin glass model~\cite{gard}.
 From this viewpoint, influence of an edge size on the typical estimates of random computational problems
 is investigated by statistical-mechanical approaches.
% In statistical-mechanical terms, the issue is also investigated.
 In fact, it has been revealed that the edge size changes the properties of
 some problems such as $K$-SAT~\cite{25,26}, $q$-coloring~\cite{qc} and min-VCs on
 $K$-uniform regular random hypergraphs~\cite{4}.
 It is also found that there exists a P/NP transition between 2-SAT and 3-SAT~\cite{22}.
%The edge size determines
%the number of spins feeling one interaction, so more than three body interaction is taken into consideration
%on hypergraphs.
%In order to study the typical case of the min-VC described by the multi-body interactions,
%we consider the problems on random $\alpha$-uniform hypergraphs.
 Here we study the typical case of the size of the min-VC, explained later, on random $\alpha$-uniform hypergraphs
 and
 focus on the relation between the replica symmetry and the performance of an approximation algorithm
 called a leaf removal algorithm.
% First, we show the theoretical analysis of min-VCs by the statistical-mechanical approaches.
% Then, we consider the leaf removal algorithm. It solves the min-VCs on sparse and tree-like graphs with high
% probability, but cannot solve them on more complex graphs. We observe the behavior of the algorithm and
% study it by a recursive analysis.
 
Let us suppose that an $\alpha$-uniform hypergraph $G=(HV,HE)$ consists of $N$ vertices $i\in HV=\{1,\cdots,N\}$ and
 (hyper)edges $(i_1,\cdots,i_\alpha)\in HE \subset HV^{\alpha}\, (i_1<\cdots<i_\alpha)$.
 We define covered vertices as a subset $HV'\subset HV$ and covered edges as
 a subset of edges connecting to at least a covered vertex.
 The vertex cover problem on the hypergraph $G$ is to find a set of the covered vertices $HV'$ by which all edges
 are covered.
 We define the cover ratio on $G$ as $|HV'|/N$ with $|HV'|$ being the
 size of the vertex cover problem.
 The min-VC on $G$ is to search a set of the covered vertices with the minimum cover ratio.
 %In order to study the typical average of the minimum cover ratio on the random hypergraphs,
 %we consider
 In the random $\alpha$-uniform hypergraph
 all the edges are set independently from all $\alpha$-tuples of vertices with probability $p$.
 The degree distribution of the graph converges to the Poisson
 distribution with the average degree $c$, which
 is given as $c=pN^{\alpha-1}/(\alpha-1)!$ for large $N$. 
% As the number of $\alpha$-tuples is $\left(\begin{smallmatrix}N\\ \alpha\end{smallmatrix}\right)$, we find
% that $p$ equals $(\alpha-1)!c/N^{\alpha-1}$ with the average degree $c$ if $N$ is sufficient large.
 In this Letter, we focus on an average of the minimum cover ratio $x_c$
 over the sparse random hypergraphs
 with the average degree $c$ being $O(1)$. 

 %Hereinafter we just call this average the average minimum-cover ratio.
 
The vertex cover problems are mapped on the lattice
 gas model~\cite{wh1,wh2,wh3} on the random hypergraphs.
% We consider that at most one gas particle can be placed
% on each vertex of a graph.
% We define that variables $\underline{\nu}=\{\nu_i\}=\{0,1\}^N$ represent
% the existence of particles on vertices $i\in HV$.
% In order to simplify our notations, we define $\nu_i$ as
 We define a variable $\nu_i$ on each vertex, representing the existence of a gas particle,
 which takes 0 if a vertex $i$ is covered and 1 if uncovered.
% We find that there exists such a vertex $i\in\{i_1,\cdots,i_\alpha\}$
% as $\nu_i=0$ iff an edge $(i_1,\cdots,i_\alpha)\in HE$ is covered.
% Instead of adding a penalty term to the Hamiltonian~\cite{zh}
 An covered edge has at least a vertex with $\nu_i=0$ in its connecting vertices.
 Thus, an indicator function for a given particle configuration $\underline{\nu}=\{\nu_i\}=\{0,1\}^N$
 is defined as
\begin{equation}
\chi(\underline{\nu})=\prod_{(i_1,\cdots,i_\alpha)\in HE}(1-\nu_{i_1}\cdots\nu_{i_\alpha}), \label{eq01}
\end{equation}
 which takes 1 if $\underline{\nu}$ is a solutions of the vertex cover problem on the hypergraph, and 0 otherwise.
% The vertex cover problem on the hypergraph $G$ is translated into the model whose Hamiltonian is
% $-\sum_i\nu_i$.
 Using the indicator function, the grand canonical partition function of the model reads
 \begin{equation}
 \Xi=\sum_{\underline{\nu}} \exp\left(\mu\sum_{i=1}^{N} \nu_i\right)\chi(\underline{\nu}), \label{eq02}
 \end{equation}
 where $\mu$ is a chemical potential and the sum is over all configurations of $\underline{\nu}$.
 In this formulation, only the solutions of the vertex cover problem contribute the partition function and
 its ground states in a large $\mu$ limit are given by the solutions of the min-VC.
 To study the typical case of min-VCs
 %thermodynamical property of random hypergraphs,
 we need to take the average over the random hypergraphs and the limit as $N\rightarrow\infty$.
 Then, the average minimum-cover ratio is represented as
 \begin{equation}
 x_c(c)=1-\lim_{\mu\rightarrow\infty}\lim_{N\rightarrow\infty}
 \frac{1}{N}\mathsf{E}\left\langle\,\sum_i\nu_i\,\right\rangle_\mu, \label{eq03}
 \end{equation}
 where $\langle\cdots\rangle_\mu$ is the grand canonical average and
 $\mathsf{E}$ is the average over the random hypergraph ensemble. % which is characterized by the average degree $c$ and the edge size $\alpha$. 
 Our aim is to obtain the theoretical estimate of the average
 minimum-cover ratio as a function of the average degree $c$.
 
%Because the average minimum-cover ratio is represented by the density
%of the model,  we need to calculate the disordered  average of the
%grand potential density $-(\mu  N)^{-1}\mathsf{E}\ln\Xi$.  While we can
%use each the replica method or the cavity method, we use the former
%method.  The replica method on finite connectivity graphs~\cite{10} are
%adapted.  To calculate the replicated system, we assume the RS ansatz,
%the replica-index  permutation invariance of order parameters. 
The average minimum-cover ratio is derived from the averaged grand
 potential density $-(\mu  N)^{-1}\mathsf{E}\ln\Xi$, which is obtained by
 using the replica method for finite connectivity graphs~\cite{10}.
 Following the standard procedure of the replica method, the original problem
 is reduced to solving a saddle-point equation of a replicated order
 parameter functional. To proceed the calculation, we assume the RS ansatz that the
 solution of the saddle-point equation has a replica symmetry.
 Introducing a local field on a vertex associated to the order
 parameter and its distribution function,
 we obtain the saddle-point equation of the distribution. 
Finally, under the RS ansatz, the average minimum-cover
 ratio is obtained as a function of the average degree $c$,
 \begin{equation}
x_c(c)=1-\left[\frac{W((\alpha-1)c)}{(\alpha-1)c}\right]^{\frac{1}{\alpha-1}}
\left(1+\frac{W((\alpha-1)c)}{\alpha}\right), \label{eq04}
 \end{equation}
 where $W(x)$ is the Lambert W function defined as $W(x)\exp(W(x))=x$.
 We call this estimate the RS solution of min-VCs. 
This solution is also obtained by an alternative cavity method~\cite{zh}. 
Although the instability of the RS solution such as the de Almeida-Thouless
 instability~\cite{at} must be examined to validate the solution, we here
 naively study an instability condition of the saddle-point equation
 against a perturbation of the local field distribution within the RS
 sector.  The analysis leads to a critical value of the average
degree $c_*=e/(\alpha-1)$
 above which the RS solution becomes unstable. 
These results, $x_c$ and $c_*$, include the case of $\alpha=2$~\cite{wh1}. 
The obtained $x_c$ gives a correct value below the critical average
degree, while a RSB solution for $x_c$ is required above it. 
% Especially in the case of $\alpha=3$, we estimate the RS solution below the critical value $c^\ast=e/2$, and
% the replica symmetric solution is no more stable in the RSB phase $c>e/2$.
% \textbf{Note that we would check the instability by some methods like
% the Almeida-Thouless condition~\cite{at} instead of adding the perturbation.}
 
\begin{figure}[!tb]
\begin{center}
 \includegraphics[trim=0 0 0 0,width=1.0\linewidth]{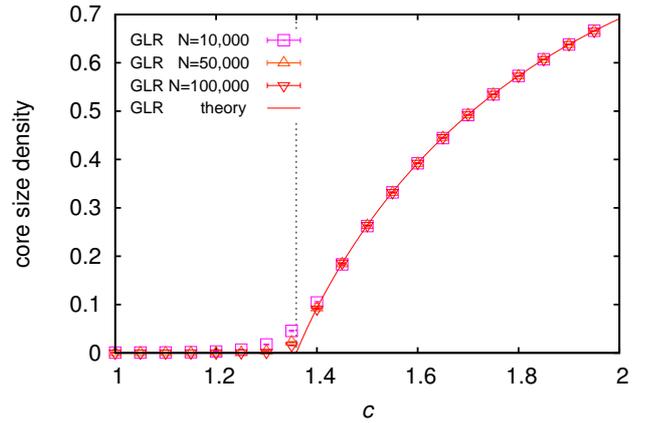}
 \caption{(Color Online). The core size density in the GLR algorithm as a function of the average degree $c$.
 Open marks are the data obtained by the GLR algorithm with the vertex size $10^4$,
 $5\times 10^4$, and $10^5$, which are taken an average over $10^4$ random hypergraphs.
 The solid line is the core size density predicted by our
 recursive analysis. The vertical dotted line represents the critical average degree $c_\ast=e/2$.}
 \label{fig01}
\end{center}
\end{figure}

% Now we estimate the average minimum-cover ratio by the
% statistical-mechanical approach, we consider  the approximation
% algorithm to search solutions of min-VCs in polynomial time.
Here we turn our attention to the estimate of $x_c$ by using an
 approximation algorithm.
 The leaf removal algorithm has been
 proposed as an approximation
 algorithm to solve a min-VC on a graph with $\alpha=2$~\cite{ks}
 and has also been applied to search for a $k$-core~\cite{xor1} %in terms
				%of the graph theory 
and a 3-XOR-SAT solution~\cite{24}. 
For a min-VC on a given graph, this algorithm consists of 
iterative steps, where 
vertices called a leaf, as well as the edges connecting to the leaves, are 
removed from the graph with covered vertices appropriately assigned to those vertices.
 This removal step makes new leaves and the algorithm continues in an iterative 
way until the leaf is empty. By this procedure, the minimum cover ratio 
is estimated correctly at least for the removed part of the graph.
 We consider the global leaf removal (GLR) algorithm~\cite{bg}, which removes
 simultaneously all the leaves found in a recursive step.
%We extend this algorithm to the hypergraph with $\alpha=3$. 
We focus on the expansion of this algorithm for the min-VC on a hypergraph with $\alpha=3$,
 while it is straightforward to extend it to that on a hypergraph with $\alpha\ge 4$.
A crucial point in our algorithm is in definition of leaf, where 
 a leaf $\{i,j,k\}\in HV^3\,(i<j<k)$ is defined as a 3-tuple of vertices
 connecting to an edge $(i,j,k)$, at least two of which the degree is one.
% Modyfing the definition of the leaf, the algorithm can be straightforward
% expanded to hypergraphs with $\alpha\ge 4$.
 The definition of the GLR algorithm is as follows:
 \begin{description}
 \item[Step 1] The initial graph $G$ is named $G^{(0)}$. Set $k=0$.
 \item[Step 2] Search all leaves from the graph $G^{(k)}$. If there is no
	    leaf, go to \textbf{Step 6}.
            %the recursive step for a removal stops.
 \item[Step 3] Remove all the leaves except for the vertices which
	    belong to more than two leaves, named bunch of
	    leaves~\cite{bg}, and remove only one of leaves in each
	    bunch. 
 \item[Step 4] Assign covered vertices to the one with the maximal degree
	    in each removed leaf from $G^{(k)}$. 
 \item[Step 5] %The $k$-th step is finished. 
The left graph is named $G^{(k+1)}$, and return to \textbf{Step 2}
 with $k$ incresed by one. 
 \item[Step 6] If there exist connected vertices
 in the left graph, assign all of them to covered vertices.
 Stop the algorithm.
\end{description}
It is proven that the result of the algorithm is independent of
 order of removal and a selection of a leaf out of a
 bunch of leaves in the removal process.
 When the recursive steps stop, the left graph consists of isolated vertices
 and a core, which is defined as 
 a set of vertices connecting to edges without leaves.
 Vertices in a bunch of leaves which are not selected for the
 removal in
 \textbf{Step 3} become isolated and the core of the
 order $O(N)$ exists in large $c$. 
% We consider the average over the random hypergraphs as well as the
% statistical-mechanical analysis  and take a limit as
% $N\rightarrow\infty$.
We note that \textbf{Step 4} can be omitted if one is interested
 only in the minimum cover ratio,
 not the covered vertices. 
% , which is
% given as an one-third of the fraction of all the removed vertices with .
Because the algorithm %cannot solve the min-VC in the core,
 covers all vertices in the core without searching the solution of the min-VC
 as shown in \textbf{Step 6},
 the existence of the core of the order
 $O(N)$ leads to overestimation %a wrong estimate
 of the average minimum-cover ratio.  
We study the core size at the end of the GLR algorithm by numerically
 performing the above-mentioned procedure for finite-size random
 hypergraphs with $\alpha=3$. 
While the computational time for the GLR algorithm is proportional to
 the number of vertices, 
% we can perform the algorithm up to typically $N=10^6$. However if we
 % generate a random  graph as the definition, 
it takes time of the order $O(N^3)$ for generating a random graph. 
To avoid it, we use %consider another strategy to  create a random hypergraph. As the expectation number of edges is $cN/\alpha$, we fix the number of edges
the microcanonical ensemble~\cite{bg} with fixing the number of edges to
the expectation number of edges $cN/3$, ignoring fluctuation of the
average degree. 
%we consider that it leads  to same results compared to generating graphs
%following the definition if $N$ is sufficient large. 
We expect that such fluctuation is irrelevant in a large size $N$
limit. 
In Fig.~\ref{fig01}, the core size density obtained by numerical 
simulations is presented as a function of the average degree $c$ up to 
the size $N=10^5$. The data averaged over $10^4$ random graphs converges 
well for large sizes and a giant core with $O(N)$ emerges above a 
certain value of $c$. 
 %If the graph is tree-like, the algorithm completely removes the graph and provides an exact estimate of the
 %minimum cover ratio. However, for a loopy graph, a part of a graph with no leaves, called the core, remains at the end of the
 %recursive steps, and the algorithm cannot solve the min-VC.
%% \begin{figure}[!tb]
%%\begin{center}
%% \includegraphics[trim=0 0 0 0,width=1.0\linewidth]{./glr_core2.eps}
%% \caption{(To be compared with new figure.)}
%%\end{center}
%%\end{figure}

We discuss the asymptotic behavior of the recursive procedure in the GLR algorithm. 
%We study the leaf removal algorithm by recursive analysis~\cite{bg}
%though we can also examine it 
%instead of solving rate equations~\cite{wh3}. 
We introduce the average fraction of the core $c_n$ and the isolated vertices $i_n$
 over random hypergraphs after $n$-th step of the algorithm, and find %~\cite{th}
 \begin{equation}
 \begin{split}
i_n&=e_{2n+1}+2e_{2n}+2ce_{2n}e_{2n-1}^2-2,\\
c_n&=e_{2n}-e_{2n+1}-2ce_{2n}e_{2n-1}^2+2ce_{2n-1}^3, \label{eq05}
 \end{split}
 \end{equation}
 where a parameter $e_n$ obeys a recursion relation
 $e_{n}=\exp(-ce_{n-1}^2)$ with the initial condition $e_{-1}=0$. 
 A detailed derivation of the formulas will be reported in a
 separate paper~\cite{th}.
By definition, the average fraction of the removed vertices $r_n$ up to
 the $n$-th step is given by $r_n=1-i_n-c_n$. 
These fractions are governed by the sequence of $e_n$ and their values
 at the end of the algorithm are determined by the asymptotic behavior of
 the recursion relation of $\{e_n\}$. 
% To estimate the fraction when algorithm stops, we take a limit as
 % $n\rightarrow\infty$ and
It is found that there exists a critical average degree $c_\ast=e/2$ for
 the recursion relation. Below the critical value, the sequence
 $\{e_n\}$ converges to the unique value $[W(2c)/(2c)]^{1/2}$ and
 consequently the core size $c_\infty$ is zero. 
Above the critical value, however, a bifurcation occurs in the recursion
 relation and the sequence has a cycle with period two. 
This type of the transition would occur above $\alpha=3$ at the critical average degree
 $c_\ast=e/(\alpha-1)$.
Because $e_{-1}=0$,
 an even term $e_{2n}$ is larger than that at one-step later, that is $e_{2n+1}$. 
We compute the limiting values $\lim_{n\rightarrow\infty}e_{2n+1}$ and
 $\lim_{n\rightarrow\infty}e_{2n}$ numerically as a function of $c$. 
The difference between them yields emergence of the core of the order of $O(N)$. 
We present the core size density obtained from the asymptotic analysis of the
 recursion relation by the solid line in Fig.~\ref{fig01}, which
 coincides with the data by numerical simulations. Thus, we confirm that
 a core percolation occurs at the critical average degree in the GLR algorithm,
 which coincides with that of the RS instability.
 From the analysis near the critical degree, it is found that 
% We perform the algorithm on random 3-uniform hypergraphs to verify the prediction.
% Fig.~\ref{fig01} shows that the numerical and analytic results of the core size are coincident.
the size of the core emerges linearly near above the critical average degree. %,
These findings, the bifurcation in the recursion relation and the core
percolation, are common in the min-VCs on random graphs
with $\alpha=2$. 
 %and monotonously increases up to 1  as the average degree $c$ increases.

As mentioned above, the GLR algorithm estimates the minimum cover
 ratio by the size of the removed part in the graph during the
 recursive procedure, which is given as $r_\infty=1-i_\infty-c_\infty$.
 % defined as the
 % deleted vertices on $G^{(\infty)}$  from the initial graph $G^{(0)}$
 % is 
 Taking one-third of $r_\infty$ and adding $c_\infty$ to
 the value, we obtain the estimate of the average minimum-cover ratio
 by the algorithm. 
Thus, we find that below the critical average degree $e/2$ the estimate $r_\infty/3$
 coincides with the RS solution Eq.~(\ref{eq04})
 estimated by the replica method.
 In contrast, the sequence $\{e_n\}$ of the algorithm does not converge to a
 unique value above the critical value and the GLR algorithm
 could not give a precise estimate of $x_c$ there. 

\begin{figure}[!tb]
\begin{center}
 \includegraphics[trim=0 0 0 0,width=1.0\linewidth]{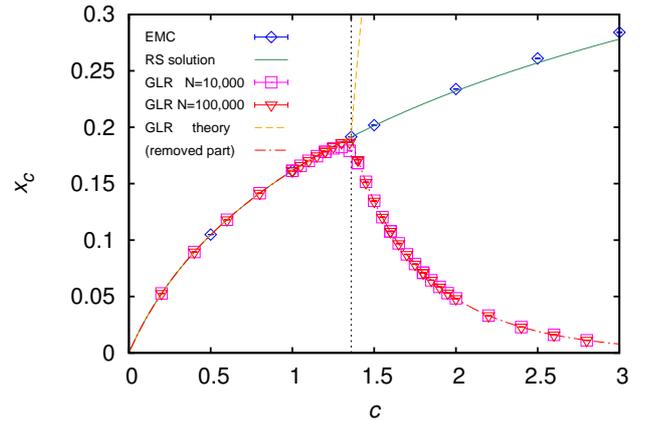}
 \caption{(Color Online). The average minimum-cover ratio on random $\alpha$-uniform
 hypergraphs with $\alpha=3$ as a function of the average degree $c$. 
 Open marks are numerical results by the exchange MC (diamonds)
 and by the GLR algorithm for $N=10^4$ (squares) and $10^5$ (triangles).
 Lines represent analytical results by the replica method (solid), by the GLR algorithm
 (dashed) and on the removed part of the graphs by the GLR algorithm (dashed-dotted).
 The vertical dotted line is the critical average degree $c_\ast=e/2$, below which
 all lines merge into a single line.
 %The vertical dashed line represents the critical average degree .
 }
 \label{fig02}
\end{center}
\end{figure}

%We show the statistical-mechanical analysis by the replica method
%and  performance of the global leaf removal algorithm. 
In order to confirm whether these analyses estimate the average
 minimum-cover ratio $x_c$ correctly, we also evaluate the min-VCs by the Markov chain Monte Carlo method.
 We use the replica exchange Monte Carlo method (EMC)~\cite{91},
 for accelerating the dynamics of the system, with 50 replicas
 in the range of the chemical potential from $-2$ to $10$. 
In our Monte Carlo simulations, the smallest cover ratio found in
 typically $2^{17}$ Monte Carlo steps is used as the estimate of $x_c$
 for each random graph, which is averaged over 800 hypergraphs randomly
 generated.  
The number of vertices of the graph is up to $N=512$. 
The average minimum-cover ratio is extrapolated from
 these numerical results for finite $N$. 
%To average the minimum cover ratio, we generate  800 random hypergraphs, and use only the smallest cover ratio appeared in  per one random graph.
% We simulate random hypergraphs whose number of vertices 
% and estimate the average minimum-cover ratio by extrapolating from these numerical results.
 Fig.~\ref{fig02} shows the obtained minimum cover ratio as a function
 of the average degree $c$.
Below the critical average degree $e/2$ where the RS solution is
 considered to be correct, 
 we observe that the MC result is consistent with those by the two
 approaches, the replica method and the GLR algorithm. 
 Above the critical value, on the other hand, the MC estimate stays
 slightly above that by the replica method and considerably deviates
 from that by the GLR algorithm. 
The former is due to the instability of the RS solution and the latter
 is the existence of the core of the order $O(N)$. 
 
%%\begin{figure}[!tb]
%%\begin{center}
%% \includegraphics[trim=0 0 0 0,width=1.0\linewidth]{./glr_xc.eps}
%% \caption{(To be compared with new figure.)}
%%\end{center}
%%\end{figure}  
To summarize, we consider the minimum vertex cover problems on random $\alpha$-uniform hypergraphs,
 and analyze them by the statistical-mechanical method and the approximation algorithm. The replica method
 estimates the average minimum-cover ratio $x_c$ as a function
 of the average degree $c$ under the replica symmetric assumption. 
 We find that there is an RS/RSB phase transition 
at the  critical average degree $c_\ast=e/(\alpha-1)$, which is well above a
 percolation threshold $c=1/(\alpha-1)$ in the random graph. 
We also perform the global leaf removal algorithm and study the
 asymptotic behavior of the recursive procedure of the algorithm,
 particularly in the case of $\alpha=3$.
 If the average degree is below the critical value which coincides with
 that in the replica theory, there is a core of the order $O(1)$
 in the remaining part of the graph, which does not
 affect the estimate of the minimum cover ratio. 
In contrast, above the critical value, the core of the order $O(N)$
 emerges, leading to a wrong estimation of
 the minimum cover ratio. 
%because the algorithm cannot identify covered vertices in the core.
% because the algorithm covers all the vertices in the core.
% We also simulate the replica exchange Monte Carlo method to observe
% the exact random graph average of the minimum cover ratio. 
Comparing the results obtained by MC simulations, we confirm that these
 estimates are correct below the critical average degree, but this is not
 the case above the critical degree. 
 These results strongly suggest that there is a close relation between
 the replica symmetry in statistical physics and the performance of the
 leaf removal algorithm even when the edge size $\alpha$ is
 larger than two. 

It is noted that this relation is not always true for all types of
 random graphs. 
For instance,  
the GLR algorithm removes no vertex on regular
 random graphs with $c \ge 2$ because no leaf is found there while, from the point of the
 statistical-mechanical view, the min-VCs on regular random 2-uniform
 graphs with degree 2 is described by the RS
 solution~\cite{4}. 
Thus, the relation depends on a type of random graphs and
 approximation algorithms. 
In addition to the leaf removal algorithm, a recent work for the min-VC
 problem with $\alpha=2$~\cite{dh} suggests that 
 linear programming algorithms,
 which are one of the most commonly used tools for solving optimization
 problems, % problems of an integer programming problem,  
have the relation discussed in the present work. % is reproduced by adding  the
 % extra constraints called a cutting-plane approach. We examine it
 % without the cutting-plane approach  on 3-uniform hypergraphs and find that
% the similar relation appears~\cite{th}.
Further study will need to establish the relation between the replica symmetry 
and the performance of numerous algorithms.

This research was supported by a Grants-in-Aid for Scientific Research
 from the MEXT, %Ministry of Education, Culture, Sports, Science and Technology,
 Japan, No. 22340109.

\end{document}